\documentclass[multphys,vecphys]{svmult}

\usepackage{makeidx}         
\usepackage{graphicx}        
\usepackage{multicol}        
\usepackage[bottom]{footmisc}
\usepackage{natbib}

\makeindex             

\newcounter{Rco}
\newcommand{\Ionst}[1]{\setcounter{Rco}{#1}\Roman{Rco}}
\newcommand{\Ion}[2]{\mbox{#1\ {\scriptsize\Ionst{#2}}}}

\newcommand{\Teff}{\mbox{$T_\mathrm{eff}$}}
\newcommand{\lsv}{\mbox{LS\,V\,$+46^{o}21$}}
\newcommand{\pnsh}{\mbox{Sh\,2$-$216}}

\begin{document}

\title*{HST Spectroscopy \\ of the Hottest White Dwarfs}


\author{Thomas Rauch
\and
        Klaus Werner}


\institute{Institut f\"ur Astronomie und Astrophysik, Sand 1, 72076 T\"ubingen, Germany
\texttt{rauch@astro.uni-tuebingen.de}
}
%
%
\maketitle

\begin{abstract}

Spectral analysis needs the observation of lines of successive ionization stages in order
to evaluate the ionization equilibrium (of a particular species) which is a sensitive indicator
for the effective temperature (\Teff). Since stars with \Teff\ as high as 100\,000\,K
have their flux maximum in the extreme ultraviolet (EUV) wavelength range and due to the high
degree of ionization, most of the metal lines are found in the ultraviolet (UV) range.
Thus, high-S/N and high-resolution UV spectra are a pre-requisite for a precise analysis.
Consequently, we employed the Faint Object Spectrograph (FOS), the Goddard High Resolution
Spectrograph (GHRS), and the Space Telescope Imaging Spectrograph (STIS) aboard the
Hubble Space Telescope (HST) in order to obtain suitable data. We present state-of-the-art
analyses of the hottest (pre-) white dwarfs by means of NLTE model atmospheres which include
the metal-line blanketing of all elements from hydrogen to nickel.

\end{abstract}

\section{INTRODUCTION}
\label{sect:introduction}

In the early eighties of the last century, 
the evolution of ``H-normal'' post-AGB stars has been quite well understood,
e.g\@. \citet{s1983} and \citet{bs1990} have presented evolutionary calculations
for these stars. At that time, neither standard evolutionary calculations nor
model atmospheres could explain observations of H-deficient post-AGB stars.

In 1979 the discovery of PG\,1159$-$035, the H-deficient prototype of the GW Vir variables, 
had shown the inadequacy of theory: the optical spectrum exhibits broad and 
shallow  absorption lines of highly ionized species, e.g\@. \Ion{He}{2} and \Ion{C}{4},
indicating \Teff\ to be much higher than 100\,000\,K. At this temperature regime, the
assumption of local thermodynamical equilibrium (LTE) is not valid and thus, 
adequate fully metal line-blanketed NLTE model-atmospheres were required -- but not available. 

In Sect.\,\ref{sect:tmap} we describe briefly our NLTE model-atmosphere
code \emph{TMAP}, which has been developed over the last two decades and has been successfully used
for the analysis of hot, compact stars. Such analyses have continuously provided
constraints for evolutionary theory and, vice versa, predictions from 
evolutionary calculations have inspired us to search for lines of unidentified
species in UV spectra 
\citep[e.g\@.][for \Ion{Ne}{7}, \Ion{F}{6}, \Ion{Ar}{7}, and \Ion{Ne}{8}, 
respectively]{wea2004, wea2005, wea2007a, wea2007b}
provided by the HST and the Far Ultraviolet Spectroscopic Explorer
(FUSE). The synergy effect of both satellites gave us the opportunity to 
precisely analyze strategic lines from the complete UV range (from the \Ion{H}{1} Lyman edge
to the optical) and to determine photospheric properties with hitherto unprecedented
accuracy. In Sect.\,\ref{sect:pg1159} and \ref{sect:lsv4621}, we give representative examples for 
our analyses of H-deficient and H-normal post-AGB stars.
\vspace{-2mm}

\section{NLTE MODEL ATMOSPHERES}
\label{sect:tmap}

We use 
\emph{TMAP}\footnote{http://astro.uni-tuebingen.de/\raisebox{.2em}{\tiny $\sim$}rauch/TMAP/TMAP.html}, 
the T\"ubingen NLTE Model Atmosphere Package
\citep{w1986, wea2003, rd2003}, for the calculation of 
plane-parallel, chemically homogeneous models 
in hydrostatic and radiative equilibrium.
\emph{TMAP} considers all elements from H to Ni \citep{r1997, r2003}.
In the analysis of \lsv\ (Sect\@. \ref{sect:lsv4621}), e.g.,
686 levels are treated in NLTE, combined with 2417 individual lines and about 
9 million iron-group lines.

\section{SPECTROSCOPY OF PG\,1159 STARS}
\label{sect:pg1159}

PG\,1159 stars are so-called ``born-again post-AGB stars'' \citep{iea1983},
i.e\@. after their departure from the asymptotic giant branch (AGB) and at
already declining luminosity, they experienced a (very) late thermal pulse 
(He-shell flash) and returned to the AGB. During the born-again phase,
the entire H-rich envelope ($10^{-4}\,\mathrm{M_\odot}$) was convectively mixed 
\citep{hea1999, aea2005}
with the intershell material ($10^{-2}\,\mathrm{M_\odot}$, located between 
He- and H-burning shells) and H is completely burned.
The direct view on intershell matter (at the surface now) allows to 
conclude on details of nuclear and mixing processes in AGB stars. This
is an important test for stellar evolutionary models \citep[cf\@.][]{wh2006}.

Our analyses of PG\,1159 stars revealed that their abundances of 
He, C, N, O, Ne, Mg, F, Si, and Ar are in line with predictions from evolutionary 
models. These models show also a Fe depletion due to n-captures within the
s-process. In three observations of PG\,1159 stars with FUSE, no iron lines
are detectable which gives a surprisingly large Fe-deficiency of 1 -- 2\,dex 
\citep{mea2002}. An inspection of STIS observations of the same objects 
\citep[e.g\@.][]{jea2007} 
shows that there is no increase of the Ni abundance and thus, it appears likely
that the s-process has converted even Ni into trans iron-group elements.
However, we do not have reliable atomic data to prove this. Other elements
show deviations from theory, e.g\@.
P appears roughly solar but the models predict a strong enhancement while
S is expected to stay solar but shows large depletion (up to 2\,dex).
For a detailed review, see \citet{wh2006}.

\begin{figure}[ht]
\centering
\includegraphics[width=\textwidth]{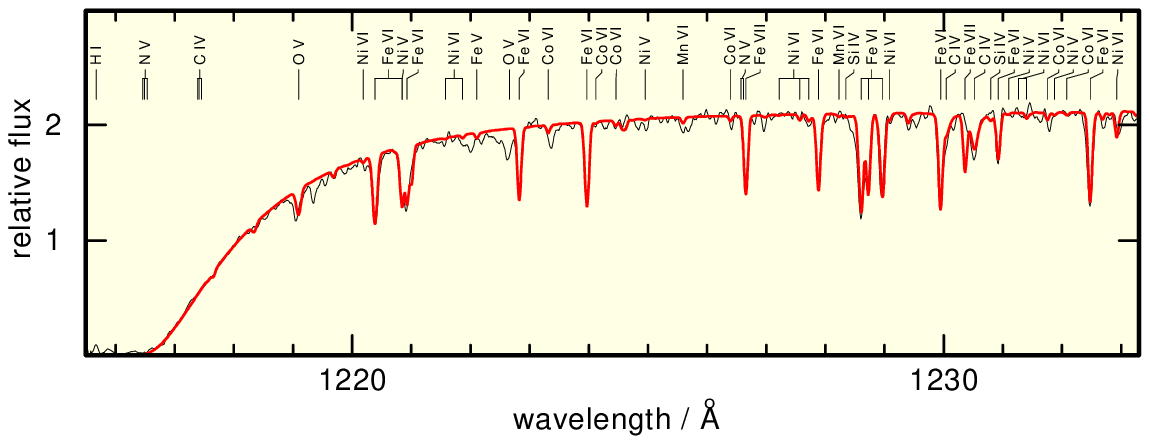}
\caption{Section of the STIS observation of \lsv\ compared to our final model. 
Identified lines are marked at the top.}
\label{fig:lsv4621}
\end{figure}

\section{SPECTROSCOPY OF \lsv}
\label{sect:lsv4621}

\lsv\ is the central star of the closest known ($d=130\,\mathrm{pc}$, $\oslash = 1.6^{o}$) 
planetary nebula \pnsh. We have observed \lsv\ with STIS (5.5\,ksec in 2000).
The STIS observation shows more than 1000 absorption features (about 10\% interstellar).
95\% of these are identified. We have calculated the most detailed \emph{TMAP} 
model-atmosphere ever \citep{rea2007} 
in order to reproduce the observed spectrum (an example is shown in Fig\@.\ref{fig:lsv4621}). 
In the STIS observation, we identified \Ion{Si}{5}
lines \citep[cf\@.][]{jea2007}, \Ion{Mg}{4} lines (for the $1^\mathrm{st}$ time in a post-AGB
star), and \Ion{Ar}{6} lines (for the $1^\mathrm{st}$ time in any star). Most of the determined
abundances are in agreement with diffusion-model predictions \citep{cea1995}.

\section{TMAP IN THE VIRTUAL OBSERVATORY}
\label{sect:summary}

The HST with its UV spectroscopic capabilities has been crucial for these analyses and the
development of \emph{TMAP}. Hopefully, the Cosmic Origins Spectrograph (COS) will continue 
the work of its very successful precursors. The comparison of our synthetic spectra with the 
observations of hot, compact stars convinced us that theory works well and we have arrived 
at a high level of sophistication.

The spectral analysis, although to be done with sufficient care, has not to remain the field
of specialists. 
Within the framework of \emph{German Astrophysical Virtual Observatory} 
(GAVO, please note that the URLs given below will change to the GAVO portal\footnote{http://www.g-vo.org/portal/} later) 
project,
we provide grids of model-atmosphere fluxes 
(\emph{TMAF}\footnote{http://astro.uni-tuebingen.de/\raisebox{.2em}{\tiny $\sim$}rauch/TMAF/TMAF.html})
as well as a WWW interface
(\emph{TMAW}\footnote{http://astro.uni-tuebingen.de/\raisebox{.2em}{\tiny $\sim$}TMAW/TMAW.shtml})
to calculate individual  \emph{TMAP} model atmospheres without detailed knowledge about theory etc.

Since the reliability of synthetic spectra is strongly dependent on the accuracy of the atomic data which
is used for their calculation, standard \emph{TMAW} calculations use predefined model atoms which are 
provided within the T\"ubingen Model-Atom Database
\emph{TMAD}\footnote{http://astro.uni-tuebingen.de/\raisebox{.2em}{\tiny $\sim$}rauch/TMAD/TMAD.html}.

While the use of the \emph{TMAF} flux grids is the easiest way for a user of the Virtual Observatory,
even individual analyses can easily be performed with appropriately adjusted model atoms.

\section*{ACKNOWLEDGMENTS}

T.R\@. is supported by the \emph{German Astrophysical Virtual Observatory} project 
of the German Federal Ministry of Education and Research (BMBF) under grant 05\,AC6VTB.

\printindex
\end{document}